\documentclass[11pt,
reprint,
 amsmath,amssymb,
 aps,
 prc,
 superscriptaddress
]{revtex4-1}

\usepackage[utf8]{inputenc}
\usepackage{graphicx}
\usepackage{dcolumn}
\usepackage{bm}
\usepackage{textcomp}
\usepackage{nicefrac}
\usepackage{multirow}
\usepackage{afterpage}
\usepackage{xspace}
\usepackage{extarrows}

\usepackage{array} 
\newcolumntype{C}[1]{>{\centering\let\newline\\\arraybackslash\hspace{0pt}}m{#1}} 
\newcolumntype{L}[1]{>{\raggedright\let\newline\\\arraybackslash\hspace{0pt}}m{#1}} 
\newcolumntype{R}[1]{>{\raggedleft\let\newline\\\arraybackslash\hspace{0pt}}m{#1}} 
\newsavebox\topalignbox
\newcolumntype{T}{>{\begin{lrbox}\topalignbox\rule{0pt}{\ht\strutbox}}c<{\end{lrbox}
\raisebox{\dimexpr-\height+\ht\strutbox\relax}{\usebox\topalignbox}}} 

\renewcommand{\a}{$\alpha$\xspace}
\newcommand{\g}{$\gamma$\xspace}
\newcommand{\eg}{\textit{e.g.}\xspace}
\newcommand{\er}{$^{166}$Er(\a,n)\xspace}
\newcommand{\ho}{$^{165}$Ho(\a,n)\xspace}
\newcommand{\etal}{\textit{et al.}\xspace}
\newcommand{\iso}[2]{$^{#1}$#2}

\newcommand{\ten}[1]{\times 10^{#1}}

\newcommand{\tal}{{\sc Talys}\xspace}

\newcommand{\sma}{{\sc Smaragd}\xspace}
\newcommand{\ecm}{E_{\text{c.m.}}}

\begin{document}

\roman{figure}

\preprint{APS/123-QED}

\title{Experimental cross sections of $^{165}$Ho($\alpha$,n)$^{168}$Tm and $^{166}$Er($\alpha$,n)$^{169}$Yb\\ for optical potential studies relevant for the astrophysical \g process}

\author{J. Glorius}
\email{glorius@iap.uni-frankfurt.de}
\affiliation{Institut f\"ur Angewandte Physik, Goethe-Universit\"at Frankfurt, Frankfurt am Main, Germany}
\affiliation{GSI Helmholtzzentrum f\"ur Schwerionenforschung, Darmstadt, Germany}
\author{K. Sonnabend}%
\affiliation{Institut f\"ur Angewandte Physik, Goethe-Universit\"at Frankfurt, Frankfurt am Main, Germany}
\author{J. G\"orres}
\affiliation{Nuclear Science Laboratory, University of Notre Dame, Notre Dame, Indiana, USA}
\author{D. Robertson}
\affiliation{Nuclear Science Laboratory, University of Notre Dame, Notre Dame, Indiana, USA}
\author{M. Kn\"orzer}%
\affiliation{Institut f\"ur Kernphysik, Technische Universit\"at Darmstadt, Darmstadt, Germany}
\author{A. Kontos}
\affiliation{National Superconducting Cyclotron Laboratory, Michigan State University, East Lansing, Michigan, USA}
\affiliation{Joint Institute for Nuclear Astrophysics, JINA}
\author{T. Rauscher}%
\affiliation{Centre for Astrophysics Research, School of Physics, Astronomy and Mathematics, University of Hertfordshire, Hatfield AL10 9AB, United Kingdom}
\affiliation{Department of Physics, University of Basel, 4056 Basel, Switzerland}
\author{R. Reifarth}%
\affiliation{Institut f\"ur Angewandte Physik, Goethe-Universit\"at Frankfurt, Frankfurt am Main, Germany}
\author{A. Sauerwein}%
\affiliation{Institut f\"ur Angewandte Physik, Goethe-Universit\"at Frankfurt, Frankfurt am Main, Germany}
\author{E. Stech}
\affiliation{Nuclear Science Laboratory, University of Notre Dame, Notre Dame, Indiana, USA}
\author{W. Tan}
\affiliation{Nuclear Science Laboratory, University of Notre Dame, Notre Dame, Indiana, USA}
\author{T. Thomas}%
\affiliation{Institut f\"ur Angewandte Physik, Goethe-Universit\"at Frankfurt, Frankfurt am Main, Germany}
\author{M. Wiescher}
\affiliation{Nuclear Science Laboratory, University of Notre Dame, Notre Dame, Indiana, USA}
\affiliation{Joint Institute for Nuclear Astrophysics, JINA}
\date{\today}

\begin{abstract}
\begin{description}
\item[Background]
Optical potentials are crucial ingredients for the prediction of nuclear reaction rates needed in simulations of the astrophysical $\gamma$ process. Associated uncertainties are particularly large for reactions involving $\alpha$ particles. This includes (\g,\a) reactions which are of special importance in the \g process.
\item[Purpose]
The measurement of ($\alpha$,n) reactions allows for an optimization of currently used \a-nucleus potentials. The reactions $^{165}$Ho($\alpha$,n) and $^{166}$Er($\alpha$,n) probe the optical model in a mass region where \g process calculations exhibit an underproduction of \textit{p} nuclei which is not yet understood.
\item[Method]
To investigate the energy-dependent cross sections of the reactions $^{165}$Ho($\alpha$,n) and $^{166}$Er($\alpha$,n) close to the reaction threshold, self-supporting metallic foils were irradiated with \a particles using the FN tandem Van de Graaff accelerator at University of Notre Dame. The induced activity was determined afterwards by monitoring the specific $\beta$-decay channels.
\item[Results]
Hauser-Feshbach predictions with a widely used global \a potential describe the data well at energies where the cross sections are almost exclusively sensitive to the \a widths. Increasing discrepancies appear towards the reaction threshold at lower energy. 
\item[Conclusions] The tested global \a potential is suitable at energies above $14$ MeV, while a modification seems necessary close to the reaction threshold. Since the \g- and neutron width show non-negligible impact on the predictions, complementary data are required to judge whether or not the discrepancies found can to be solely assigned to the \a width.
\end{description}
\end{abstract}

\maketitle


\section{\label{sec:intro}Introduction}

The nucleosynthesis of elements heavier than iron is dominated by neutron capture processes \cite{kaep11,arno08,ctt}. However, there are 35 proton-rich isotopes, which cannot be synthesized by neutron capture reactions of the \textit{s} and \textit{r} processes \cite{raus13,Arno03}. The bulk of these so-called \textit{p} nuclei are believed to be produced in a series of photodisintegration reactions, referred to as the \g process \cite{raus13,woohow}, which is found in the outer layers of massive stars during their explosion in a core-collapse supernova \cite{woohow,rayet95,Raus02} or in explosions of White Dwarfs as type Ia supernovae \cite{howmey,kusakabe,travaglio}. At temperatures between 2 and 3 GK, an initially present heavy seed distribution is first altered by (\g,n) reactions and thus proton-rich nuclei are created. After several neutron emissions, (\g,p) and (\g,\a) reactions start to compete, deflecting the reaction flux. The whole process lasts only a few seconds, before the environment becomes too cool to permit photodisintegration. Eventually, $\beta$-decay chains populate the next stable isobar encountered for a given mass number. Detailed reaction network calculations (e.g., \cite{rapp06,Raus02}) have to consider several hundred reactions for which the reaction rates under the given conditions have to be known. Since many potentially important reactions involve unstable target nuclei, measurements are challenging or even impossible and the experimental database of relevant reactions for the \g process is far from being complete \cite{raus13,kadonis}. 

\begin{figure}[t]
\includegraphics[width=\columnwidth]{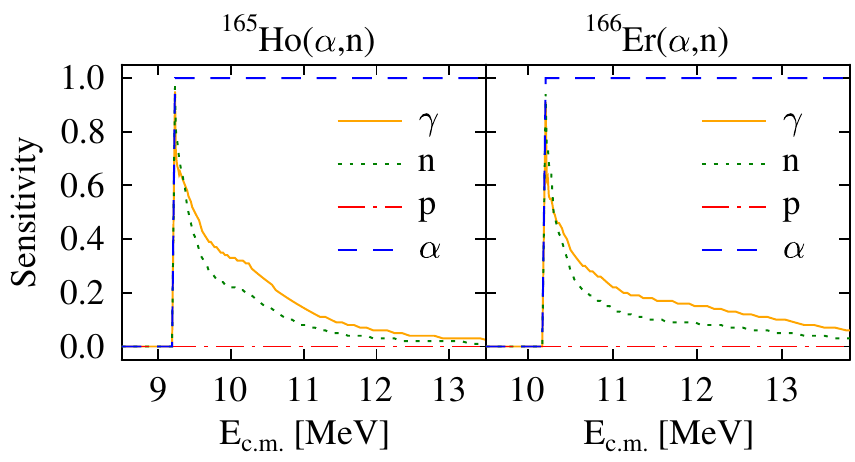}
\caption{(Color online) Sensitivities of the investigated laboratory cross sections to four different channel widths versus center-of-mass energy \cite{raus12a}. See text for details.}
\label{fig:one}
\end{figure}

\begin{figure*}[t]
\includegraphics[width=0.95\textwidth]{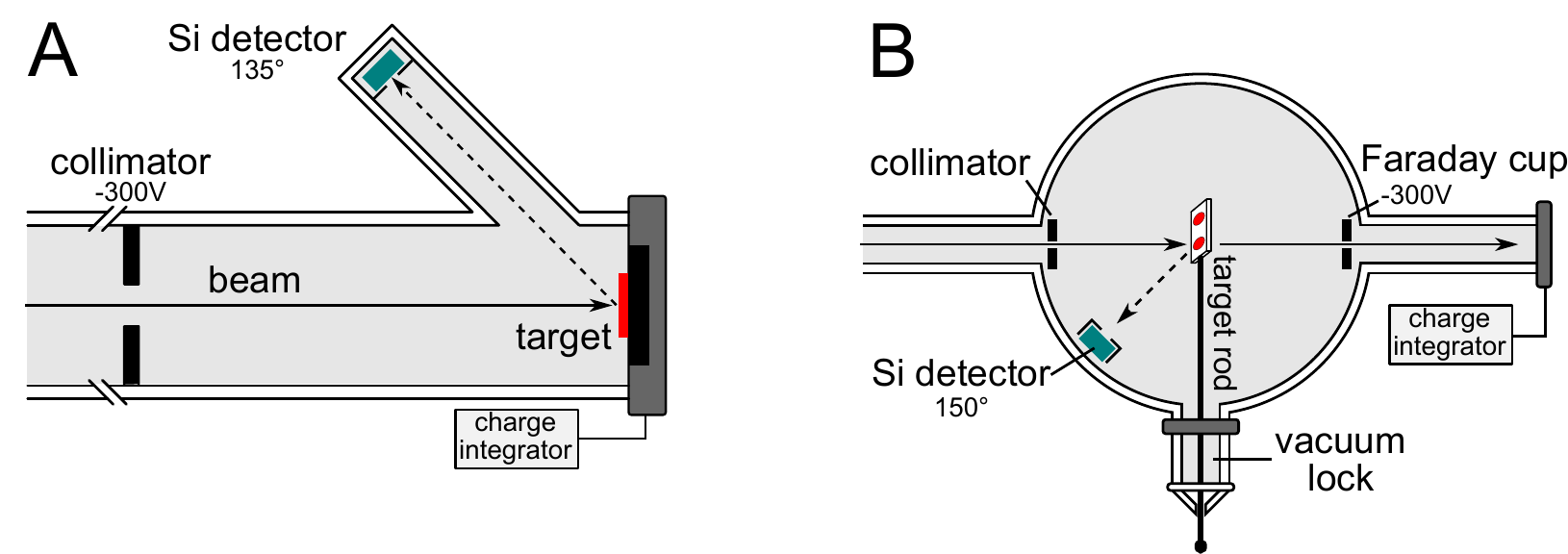}
\caption{(Color online) Sketches of the setups used during the campaign. Setup {\it A} (left side) was used in the first period, while the improved setup {\it B} was used during the two subsequent periods. Both setups feature a beam current measurement, a suppression voltage and a Si detector. The vacuum lock integrated in setup {\it B} allows a rapid exchange of samples. Note that the setups are not drawn to scale.}
\label{fig:chamber}
\end{figure*}

Consequently, the bulk of reaction rates has to be predicted by theory making use of the Hauser-Feshbach approach \cite{Haus52,Raus11}. A crucial input parameter of this model are optical model potentials, which describe the interaction between nuclei and light nuclear particles. Of particular importance for the \g process is the \a-nucleus potential, being a crucial ingredient for the prediction of (\g,\a) reaction rates, which determine the reaction flux in the heavy mass region. Large deviations are present between predicted and measured \a-induced cross section data in the energy region of relevance, see e.g. \cite{raus13,Somo98,gyur06}. 

Usually, \a-nucleus potentials are derived from or adjusted to elastic \a-scattering data, see e.g.\ \cite{avri09,McFa66}. However, for energies well below the Coulomb barrier the scattering process is strongly dominated by the Coulomb interaction and a measurement of the nuclear component becomes unfeasible. Therefore, such measurements are not very sensitive to the imaginary part of the optical potential, which accounts for nuclear absorption of the incoming particles \cite{Raus11,fuel01}. Hence, to improve the predictive power of the \a-nucleus potential for energies and reaction channels relevant for astrophysics, an extended database of suitable reaction cross sections has to be produced, preferably for intermediate and heavy nuclear masses. Eventually, such a database will be the starting point for a global adjustment of existing \a-nucleus potentials, leading to improved predictions of astrophysical reaction rates involving \a particles.

The cross sections of (\a,n) reactions well above the threshold can be used to constrain the \a width because it remains much smaller than the neutron width due to the Coulomb barrier.
The absolute value of the sensitivity, as defined in Ref.\ \cite{raus12a}, of the laboratory cross section is plotted in Fig.\ \ref{fig:one} for the reactions $^{165}$Ho(\a,n) and $^{166}$Er(\a,n). Except very close to the reaction threshold, the cross sections depend nearly exclusively on the \a width, a quantity directly derived from the \a-nucleus potential. Consequently, these cross section measurements can be used to directly probe the predictive power of different \a potentials close to the astrophysically relevant energy range. For the (\a,\g) reactions on $^{165}$Ho and $^{166}$Er the Gamow window at 2 GK, which is the astrophysically relevant temperature for heavy $p$ nuclei, is located below 9 MeV \cite{Raus10}.

Within this work the reaction cross sections of $^{165}$Ho(\a,n) and $^{166}$Er(\a,n) were measured to investigate the reliability of \a-nucleus potentials in a mass region where \g process studies show a significant underproduction of \textit{p} nuclei \cite{Raus02,rapp06,raus13}. The experiments were carried out at the Nuclear Science Laboratory (NSL) of the University of Notre Dame, Indiana, USA using the activation method. In a first step, the material was irradiated with \a particles and, in the following, the induced activity was determined by \g-ray spectroscopy of the samples as well as using X-ray spectroscopy, as introduced by \cite{kisslett}.

\section{\label{sec:exp}Experimental Details}

\begin{figure*}[t]
\includegraphics[width=16.5cm]{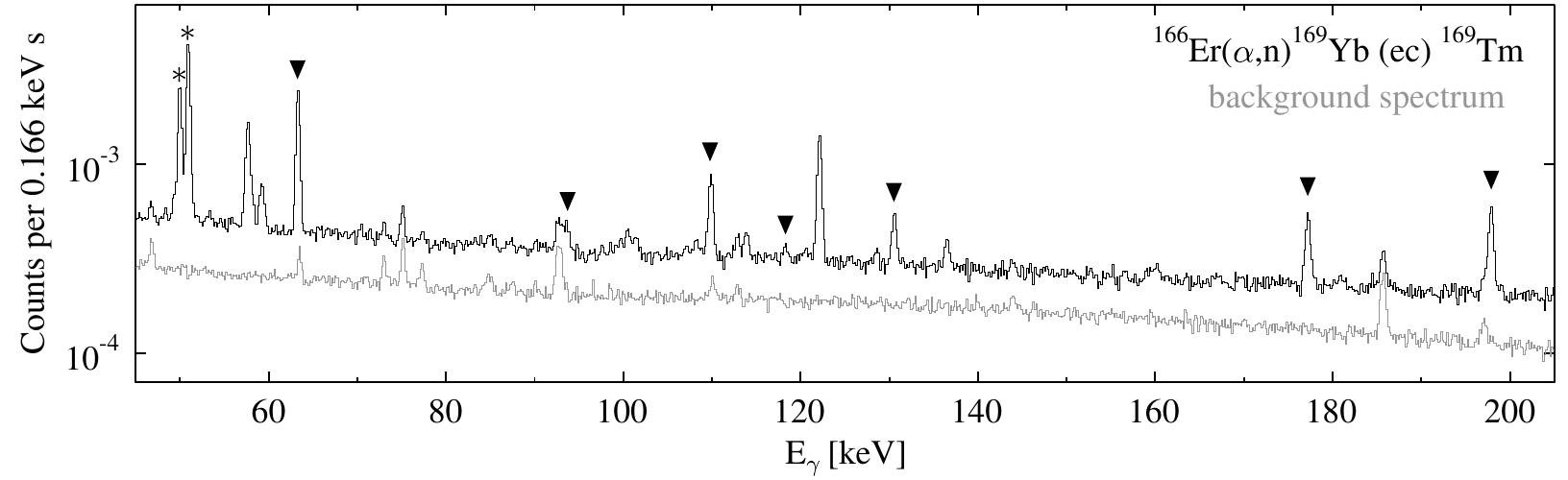}
\caption{Spectrum emitted by an erbium sample (black) after irradiation at $E_{\alpha} = 13.5$ MeV compared to the corresponding background spectrum (gray). The X-ray ($\ast$) and \g-ray ($\blacktriangledown$) transitions of \iso{169}{Tm} used for cross section analysis are marked. In order to ensure comparability, both spectra are shown in counts per $0.166$ keV s. The sample spectrum was taken for $14$ days while the background was measured for one month.}
\label{fig:spec}
\end{figure*}

The FN Tandem Van de Graaff accelerator at NSL provides a maximum terminal voltage of $11$ MV and beam currents up to $400$ nA using the duoplasmatron helium ion source. In total $20$ cross section values have been determined during the experimental campaign, which was divided in three periods. The irradiations of the first period were performed using setup $A$ (left side of Fig.\ \ref{fig:chamber}), afterwards setup $B$ (right side of Fig.\ \ref{fig:chamber}) was used for the remaining measurements. In setup {\it A} the samples are mounted on the beam stop directly and the whole chamber is designed as a Faraday cup to be able to reliably measure the beam current. In contrast, in setup {\it B} the Faraday cup is located $50$ cm behind the target position and the samples are mounted on an insulated rod providing two target positions. In both cases an electron suppression voltage is applied to an aperture in front of the beam stop to avoid corruption of the beam current measurement. Furthermore, a silicon surface barrier detector is installed at $135$\textdegree \xspace and $150$\textdegree, respectively, to monitor the target stability during irradiation by means of Rutherford Backscattering Spectrometry (RBS). Additionally, a vacuum lock allows a quick exchange of samples representing the main improvement of setup {\it B} with respect to setup {\it A}.

\begin{figure}[b]
\includegraphics[width=8.6cm]{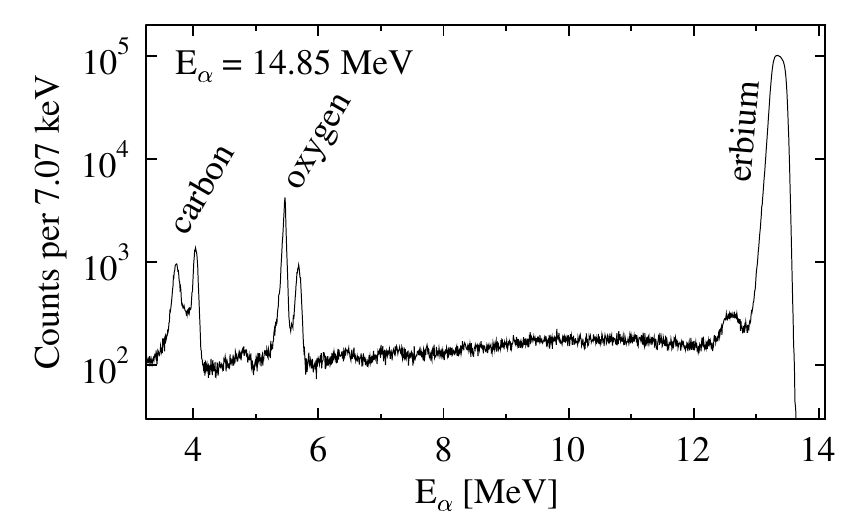}
\caption{RBS spectrum of \a particles scattered off an erbium sample. The plateau above 13 MeV corresponds to Rutherford scattering on erbium atoms. Its width is used to calculate the thickness of the sample. The small structure at the low energy tail of the erbium plateau is caused by a low amount of heavy impurities in the sample and does not disturb the analysis. At lower energies the scattering off the surface layers consisting of oxygen and carbon can be observed.}
\label{fig:RBS}
\end{figure}

Naturally composed materials of high purity ($99.99\%$) were used for the experiments and the samples were prepared by the Oak Ridge National Laboratory as self-supporting metallic foils of about 800 $\mu$g\,cm$^{-2}$ in thickness. During the activation runs, no deterioration of the sample material was observed. Each sample was used at a different beam energy and the thicknesses were extracted from the RBS spectra taken during irradiation (see Fig.\ \ref{fig:RBS}). Thus, the thicknesses extracted correspond to the average thicknesses at the location on the sample corresponding to the beam spot. The resulting values are included in Tables \ref{tab:er_xs} and \ref{tab:ho_xs}. According to the Oak Ridge National Laboratory, the variations of the thicknesses between different samples originate from the rolling procedure employed to obtain the desired thicknesses. 

After irradiation the samples were placed in front of a high-purity germanium (HPGe) detector to determine the number of induced reactions $R$. In order to achieve a high detection efficiency, the spectroscopy of decays was performed in very close sample-to-detector geometry. This can lead to strong coincidence summing caused, e.g., by \g-ray cascades \cite{Semk90}. Such effects were taken into account by calibrating the activity of a highly active sample at a large distance ($\geq 15$ cm). In such a geometry, summing effects can safely be neglected. Afterwards, the calibrated sample can be used to determine the absolute efficiency of all detectors employed in close geometry. This procedure has to be done for each decay separately to account for the specific summing of \g-ray cascades and X-rays.

Three coaxial and two planar detectors were used for spectroscopy. Each setup included a passive lead and copper shielding and supported several well defined sample positions. A cross check of all setups was carried out by multiple measurement of samples at different setups. These redundant data always agreed within the uncertainties and the respective weighted mean is given as the final result in Tables \ref{tab:er_xs} and \ref{tab:ho_xs}. An exemplary spectrum taken with a planar Low-Energy Photon Spectrometer (LEPS) is depicted in Fig.\ \ref{fig:spec}. The energy resolution of $0.5$ keV and the low background radiation level of about $10^{-3}$ counts\,keV$^{-1}$s$^{-1}$ achieved in this case allows the detection of photons in the energy range from $10$ keV to $300$ keV even if only low reaction yields are expected or weak \g transitions are involved.

\renewcommand{\arraystretch}{1.1}
\begin{table}[b]
\centering
\caption{The decay properties of the reaction products. All listed values were used for cross section evaluation.}
	\begin{tabular}{c C{0.1cm} R{1.5cm} C{0.1cm} C{0.1cm} l} \hline
decay & & \multicolumn{2}{c}{\g or X-ray} & & \multicolumn{1}{c}{\g or X-ray}\\ 
half-life $\&$ branching & & \multicolumn{2}{c}{energy [keV]} & & \multicolumn{1}{c}{intensity}\\
\hline \hline
									& & $49.77$ & & & $0.525(12)$\\
									& & $50.74$ & & & $0.916(21)$\\
									& & $63.12$  & & & $0.4362(23)$\\
\multirow{2}{*}{\iso{169}{Yb} (ec) \cite{169yb}}	& & $93.61$  & & & $0.0258(2)$\\ 
\multirow{2}{*}{$t_{\nicefrac{1}{2}} = 32.018(5)$ d} & & $109.78$ & & & $0.1739(9)$ \\
\multirow{2}{*}{$I_{\text{ec}}=100 \%$}	& & $118.19$ & & & $0.0187(1)$ \\
    								& & $130.52$ & & & $0.1138(5)$ \\
   									& & $177.21$ & & & $0.2228(11)$ \\
   									& & $197.96$ & & & $0.3593(12)$ \\
   									& & $307.73$ & & & $0.1005(5)$ \\ \hline
									& & $48.22$ & & & $0.291(8)$\\
\multirow{2}{*}{\iso{168}{Tm} (ec) \cite{168tm}}& & $49.13$ & & & $0.511(13)$\\
\multirow{2}{*}{$t_{\nicefrac{1}{2}} = 93.1(2)$ d}& & $79.80$ & & & $0.1081(39)$ \\
\multirow{2}{*}{$I_{\text{ec}}=99.990(7)\%$}& & $184.29$ & & & $0.1791(56)$ \\
									& & $198.25$ & & & $0.538(16)$ \\
									& & $447.51$ & & & $0.2367(71)$ \\ \hline
		\end{tabular}
\label{tab:decay}
\end{table}
\renewcommand{\arraystretch}{1}

To an unknown extent, the reaction \er populates an isomeric state in $^{169}$Yb, which decays to the ground state by internal transition exclusively. The altered activity in the decay channel of the ground state caused by this delayed feeding can be neglected for the evaluation, since the half-life of the isomer ($46$ s) \cite{169yb} is small compared to the half-life of the ground state ($32$ d) \cite{169yb}.

The ground state decay of \iso{169}{Yb} is followed by several \g transitions and X-ray emissions, of which the ten lines, listed in Table \ref{tab:decay}, have been observed and were used to extract cross section values. In order to use an X-ray occurrence for cross section analysis, one has to exclude contributions of other Tm isotopes. The reaction channels inducing the decays of \iso{165,167,168}{Yb} can be excluded because of a short half-life and a negligible isotopic abundance. The remaining relevant channel, \iso{162}{Er}($\alpha$,$\gamma$)\iso{166}{Yb}, features a \g decay, which should have been clearly visible in the spectra, if its X-ray contribution was significant. Thus, no interfering X-ray emissions originating from \a-induced channels were present. However, such considerations were not necessary in the case of the mono-isotopic \iso{165}{Ho} for which the analysis involved six \g-ray and X-ray energies.

\section{Data analysis}

\begin{figure}[t]
\includegraphics[width=8.6cm]{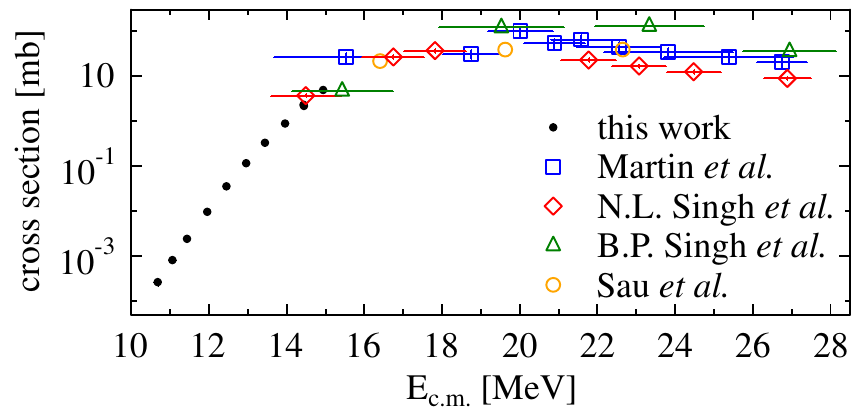}
\caption{(Color online) Comparison of the different experimental data sets for the cross section of the $^{165}$Ho(\a,n) reaction. Reasonable agreement can be observed in the overlapping region around $15$ MeV.}
\label{fig:ho_old}
\end{figure}

The cross section values presented in the next section are calculated from the number of induced (\a,n) reactions $R$, the number of \a particles $N_{\alpha}$ incident on the target and the number of target atoms per area $\mu$
\begin{equation}
\sigma = \frac{R}{N_{\alpha}~ \mu}\quad.
\end{equation}
The number of beam particles $N_{\alpha}$ impinging on the sample is measured in terms of collected charge at the Faraday cup. While $N_{\alpha}$ can be measured directly, the number of reactions $R$ and the sample thickness $\mu$ have to be derived from the spectra taken with the HPGe detectors and the Si detectors, respectively.

$R$ is directly proportional to the number of detected \g-rays or X-rays $N_{\gamma ,\text{x}}$ of the respective decay. Taking into account the dead time fraction $\tau_{\text{dead}}$, detection efficiency $\epsilon$, \g-ray or X-ray intensity $I_{\gamma,\text{x}}$ and the decay of reaction products during and after the irradiation represented by $\tau_1 \tau_2 \tau_3$, the equation can be written as
\begin{equation}
R = \frac{N_{\gamma ,\text{x}}}{\epsilon ~ I_{\gamma ,\text{x}} ~ \tau_{\text{dead}} ~ \tau_1 \tau_2 \tau_3}\quad.
\end{equation}
The correction factors $\tau_1 \tau_2 \tau_3$ are derived by applying the decay law to the periods of irradiation $\Delta t_1$, waiting time $\Delta t_2$ and spectroscopy $\Delta t_3$ and represent a link between the number of decays during spectroscopy $\Delta N$ and the total number of reactions $R$ that occurred throughout the irradiation,
\begin{equation}
\tau_1 \tau_2 \tau_3 = \frac{\Delta N}{R}.
\end{equation}
The factor $\tau_1$ accounts for decays during irradiation and apart from the decay constant $\lambda$, depends on fluctuations of the beam current $I_i$, which is measured in short time intervals $\Delta t = 10$ s,
\begin{equation}
\tau_1 =
		\frac{
			1-e^{-\lambda \Delta t }
				}{
					\lambda
				}
		\frac{
			\sum_{i=1}^{n}I_i ~ e^{-\lambda \left( 
													\Delta t_{1} - i\Delta t 													\right)}
				}{
			\sum_{i=1}^{n}I_i \Delta t
			}
		.
\end{equation}
The factors $\tau_2$ and $\tau_3$ can be derived from the decay law directly 
\begin{eqnarray}
\tau_2 & = & e^{-\lambda \Delta t_2}\\
\tau_3 & = & 1 - e^{-\lambda \Delta t_3},
\end{eqnarray}
and represent the fraction of nuclei that survive the waiting time and decay during spectroscopy, respectively.

To obtain a measure for the thickness $\mu$ of each sample, the RBS spectra taken during irradiation have been evaluated. An exemplary spectrum for erbium is depicted in Fig.\ \ref{fig:RBS}. The plateau emerging at high energies is formed by \a particles backscattered off erbium atoms in the sample. The width of this structure represents the maximum energy loss of the particles and is converted to a material thickness by making use of tabulated stopping power values \cite{srim}.

In several cases samples have been counted multiple times using different detector setups yielding multiple cross section values. These values were averaged by a weighted mean procedure. The mean values are given as final results in Tables \ref{tab:er_xs} and \ref{tab:ho_xs}.
\begin{center}
\renewcommand{\arraystretch}{1.5}
\begin{table*}[ht]
\centering
	\caption{Cross section of the reaction \er at effective center-of-mass energies $\ecm^{\text{eff}}$. Total uncertainties are given as absolute values, the varying statistical component $\Delta_{\text{stat}}$ is listed separately, while the systematic component is constant at $\Delta_{\text{syst}} = 5.1\%$. Additionally, the laboratory energy $E_{\alpha}$, the energy loss $E_{\text{loss}}$ and the areal particle density $d_{\text{RBS}}$ of the respective sample is given.}
	\begin{tabular}[c]{C{1cm} C{.6cm} C{1.1cm} C{.6cm} L{1.1cm} C{.5cm} C{1.6cm} C{.4cm} L{3cm} C{.3cm} R{1cm}}
	\hline
	\multicolumn{2}{l}{$E_{\alpha}$ [MeV]} & \multicolumn{2}{l}{$E_{\text{loss}}$ [keV]} & \multicolumn{2}{l}{$d_{\text{\tiny RBS}}$ [$\frac{\mu\text{g}}{\text{cm}^2}$]} & \multicolumn{1}{c}{$\ecm^{\text{eff}}$ [MeV]}  & &	\multicolumn{1}{c}{$\sigma$ [mb]} & & $\Delta_{\text{stat}}$ \\ \hline \hline
	$11.40$ & & $147\pm 12$ & & $860\pm 44$ & & $11.07\pm 0.02$ & &	$(3.24\pm 0.29)\ten{-4}$ & & $7.4\%$\\
	$11.75$ & & $139\pm 12$ & & $828\pm 42$ & & $11.41\pm 0.02$ & &	$(1.25\pm 0.41)\ten{-3}$ & & $32.4\%$\\
	$12.13$ & & $137\pm 12$ & & $827\pm 42$ & & $11.78\pm 0.02$ & &	$(2.96\pm 0.46)\ten{-3}$ & & $14.6\%$\\
	$12.50$ & & $126\pm 11$ & & $783\pm 40$ & & $12.15\pm 0.02$ & &	$(9.54\pm 0.61)\ten{-3}$ & & $3.9\%$\\
	$13.00$ & & $137\pm 12$ & & $865\pm 44$ & & $12.63\pm 0.02$ & &	$(3.23\pm 0.24)\ten{-2}$ & & $5.4\%$\\
	$13.50$ & & $133\pm 11$ & & $858\pm 44$ & & $13.12\pm 0.02$ & &	$(9.97\pm 0.57)\ten{-2}$ & & $2.6\%$\\
	$14.00$ & & $115\pm 11$ & & $755\pm 39$ & & $13.62\pm 0.02$ & &	$(2.91\pm 0.16)\ten{-1}$ & & $2.3\%$\\
	$14.85$ & & $120\pm 11$ & & $823\pm 39$ & & $14.44\pm 0.02$ & &	$(1.57\pm 0.08)$  		 & & $1.1\%$\\
	$15.00$ & & $114\pm 11$ & & $787\pm 40$ & & $14.59\pm 0.02$ & &	$(1.87\pm 0.10)$ 		 & & $1.8\%$\\ \hline
	\end{tabular}
\label{tab:er_xs}
\end{table*}
\begin{table*}[ht]
\centering
	\caption{Cross section of the reaction \ho at effective center-of-mass energies $\ecm^{\text{eff}}$. Total uncertainties are given as absolute values, the varying statistical component $\Delta_{\text{stat}}$ is listed separately, while the systematic component is constant at $\Delta_{\text{syst}} = 5.1\%$. Additionally, the laboratory energy $E_{\alpha}$, the energy loss $E_{\text{loss}}$ and the areal particle density $d_{\text{RBS}}$ of the respective sample is given.}
	\begin{tabular}[c]{C{1cm} C{.6cm} C{1.1cm} C{.6cm} L{1.1cm} C{.5cm} C{1.6cm} C{.4cm} L{3cm} C{.3cm} R{1cm}}
	\hline
	\multicolumn{2}{l}{$E_{\alpha}$ [MeV]} & \multicolumn{2}{l}{$E_{\text{loss}}$ [keV]} & \multicolumn{2}{l}{$d_{\text{\tiny RBS}}$ [$\frac{\mu\text{g}}{\text{cm}^2}$]} & \multicolumn{1}{c}{$\ecm^{\text{eff}}$ [MeV]}  & &	\multicolumn{1}{c}{$\sigma$ [mb]} & & $\Delta_{\text{stat}}$ \\ \hline \hline
	$11.01$ & & $152 \pm 13$ & & $741 \pm 38$ & & $10.68 \pm 0.02$ & & $(2.66 \pm 0.50)\ten{-4}$ & & $17.9\%$\\
	$11.39$ & & $152 \pm 13$ & & $879 \pm 45$ & & $11.06 \pm 0.02$ & & $(8.20 \pm 1.17)\ten{-4}$ & & $13.4\%$\\
	$11.77$ & & $142 \pm 12$ & & $834 \pm 43$ & & $11.43 \pm 0.02$ & & $(2.43 \pm 0.29)\ten{-3}$ & & $10.9\%$\\
	$12.30$ & & $129 \pm 11$ & & $777 \pm 40$ & & $11.95 \pm 0.02$ & & $(9.68 \pm 0.70)\ten{-3}$ & & $5.0\%$\\
	$12.80$ & & $113 \pm 11$ & & $700 \pm 36$ & & $12.45 \pm 0.02$ & & $(3.57 \pm 0.23)\ten{-2}$ & & $4.1\%$\\
	$13.33$ & & $125 \pm 11$ & & $792 \pm 40$ & & $12.95 \pm 0.02$ & & $(1.16 \pm 0.07)\ten{-1}$ & & $3.2\%$\\
	$13.83$ & & $123 \pm 11$ & & $798 \pm 41$ & & $13.44 \pm 0.02$ & & $(3.31 \pm 0.20)\ten{-1}$ & & $3.1\%$\\
	$14.35$ & & $120 \pm 11$ & & $797 \pm 41$ & & $13.96 \pm 0.02$ & & $(8.87 \pm 0.52)\ten{-1}$ & & $2.9\%$\\
	$14.85$ & & $132 \pm 11$ & & $895 \pm 45$ & & $14.43 \pm 0.02$ & & $(2.32 \pm 0.17)$ 	    & & $5.0\%$\\ 
	$14.85$ & & $114 \pm 11$ & & $772 \pm 39$ & & $14.44 \pm 0.02$ & & $(2.18 \pm 0.14)$ 		& & $4.1\%$\\
	$15.35$ & & $107 \pm 10$ & & $741 \pm 38$ & & $14.94 \pm 0.02$ & & $(4.92 \pm 0.29)$ 		& & $3.1\%$\\	\hline
	\end{tabular}
\label{tab:ho_xs}
\end{table*}
\renewcommand{\arraystretch}{1}
\end{center}
For each activation run an effective energy $\ecm^{\text{eff}}$ is calculated from the laboratory energy $E_\alpha$ and the energy loss $E_{\text{loss}}$ of the incident \a particles going through the sample. The relative energy dependence of the cross section $\sigma(E)$ is known in terms of center-of-mass energies $\ecm$ from the measurement and can be fitted in short intervals. With the help of this fit the effective activation energy $\ecm^{\text{eff}}$ is deduced for each sample by splitting the energy-loss interval:
\begin{equation}
\int_{\ecm - E_{\text{loss}}}^{\ecm^{\text{eff}}} \sigma (E) ~ dE = \int_{\ecm^{\text{eff}}}^{\ecm} \sigma (E) ~ dE.
\end{equation}
This integral-matching procedure is applied iteratively using repeated fits of $\sigma(E)$ to the updated energies until no change of $\ecm^{\text{eff}}$ is observed anymore.

As given in Tables \ref{tab:er_xs} and \ref{tab:ho_xs}, the final uncertainties of the cross section values are divided in statistical and systematic parts. The systematic component is dominated by the detection efficiency and the stopping power values used for the RBS analysis. It is constant at $5.1\%$ for all measurements. In contrast, the statistical component varies strongly (from $1.1\%$ up to $32.4\%$) because it is determined by the number of counts during spectroscopy. Low statistical uncertainties are achieved in cases where the cross section is deduced from the average of many transitions and several independent measurements at high count rates. If large background contributions had to be taken into account, subtraction of the latter resulted in large uncertainties. The uncertainty of the beam energy measurement $\Delta E / E = 10^{-3}$ and the precision of the energy-loss calculation $\Delta E_{\text{loss}}$ determine the final uncertainties of the effective energies given in the results. In all present cases, this uncertainty was determined to be lower than $20$ keV. 

\section{Results and discussion}

\begin{figure*}[!ht]
\centering
\includegraphics[width=16.7cm]{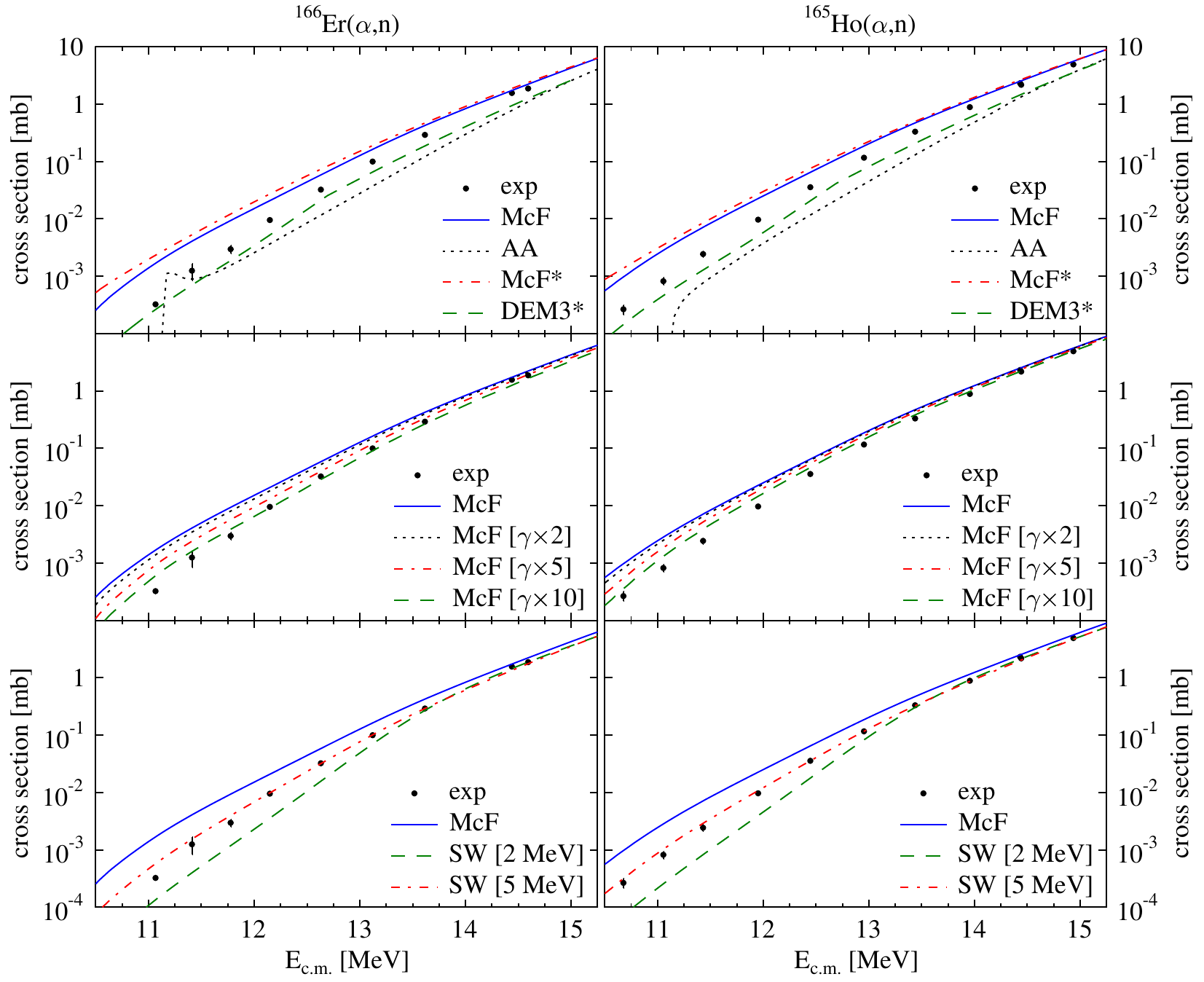}
\caption{(Color online) Experimental cross sections of the reactions $^{166}$Er(\a,n) (left side) and $^{165}$Ho(\a,n) (right side) compared to predictions of the code \sma using different input parameters. Upper panels: different global \a-nucleus potentials; also shown are results obtained with \tal (labels including asterisks). Middle panels: McF potential and scaled \g-transmission coefficients. Lower panels: predictions based on modifications of the McF potential. See text for further discussion.}
\label{fig:xs}
\end{figure*}

The final cross sections, laboratory, and effective center-of-mass energies as well as sample thicknesses are given in Table \ref{tab:er_xs} for the reaction $^{166}$Er(\a,n) and in Table \ref{tab:ho_xs} for the reaction $^{165}$Ho(\a,n), respectively. In the latter case, former cross section measurements were performed mostly at higher energies \cite{sing95,sing92,Mart66,Sau68}. These data are shown in Fig.\ \ref{fig:ho_old} in comparison to the newly derived values. In the overlapping energy range, reasonable agreement is found between the different data sets.

For both reactions, Fig.\ \ref{fig:xs} shows a comparison of the measured cross sections to values predicted by Hauser-Feshbach calculations using different assumptions for the reaction channel widths. In the upper panels, results from two Hauser-Feshbach codes are given, obtained with several optical \a-nucleus potentials and using the default input settings for each code otherwise. The code \tal \cite{taly12,koni05a} was used with the global \a-nucleus potentials of McFadden and Satchler (label McF*) \cite{McFa66} and Demetriou \etal (label DEM3*) \cite{Deme02}. Optical \a-nucleus potentials by McFadden and Satchler (label McF) \cite{McFa66} and Avrigeanu \etal (label AA) \cite{avri10} were tested with the code \sma \cite{smar12}. All further calculations (middle and lower panels) were also performed with \sma.

In order to correctly interpret the comparison between predictions and data it is necessary to recall the cross section sensitivities introduced in Fig.\ \ref{fig:one}. At the upper end of the investigated energy range, the cross sections are mainly depending on the $\alpha$ width whereas they are increasingly sensitive to $\gamma$ and neutron widths with decreasing energy. For both reactions, the McF potential reproduces the data at the higher energies well. The calculations with the other two potentials underestimate the cross sections by almost a factor of two (upper panels in Fig.\ \ref{fig:xs}). This indicates that the \a width is best described with the McF potential.

Towards lower energies, the \g- and neutron widths play an increasingly important role. This complicates the interpretation of the data because the prediction of three different widths is involved and it is not possible anymore to uniquely identify the source of the discrepancies between data and predictions. A comparison between the \tal and \sma results with the McF potential shows that the \g- and neutron widths have to be quite similar in both calculations, with only a comparatively small difference in energy dependence. Both results exhibit a different energy dependence than found in the data, with the \tal prediction being further away from the data towards lower energy. On the other hand, the \tal calculation using the DEM3* potential reproduces the energy dependence much better. This is not necessarily due to a better description of the \a width, however, but rather caused by a fortuitous cancellation of the energy dependences of all three widths. Was the same potential used with the set of widths implemented in \sma, the resulting cross sections were even lower at the lowest energies.

The divergence between calculations and data could be remedied by using different \g- and/or neutron widths. For example, scaling the \g width by a factor of ten would bring the predictions in line with the measurements for both reactions, as shown in the middle panels of Fig.\ \ref{fig:xs}. Alternatively, the neutron width could be divided by the same factor to achieve a similar result. A combination of both variations, but by smaller factors, may be a more realistic solution.

Previous work \cite{Saue11,raus12,nett13} has introduced a modification of the McF \a-nucleus potential by using an energy-dependent depth $W(E)$ of the imaginary part. It is given by a Fermi-type function
\begin{equation}
W(E) \propto \left( 1+e^{f\left(E \right)/a}\right)^{-1} \quad,
\end{equation}
where the diffuseness parameter $a$ is a constant adjusted for each target nucleus. The lower panels of Fig.\ \ref{fig:xs} show results obtained with this modified potential (label SW) for two values of $a$. At the upper end of the measured energy range, the depths of the imaginary parts approach the one of the McF potential and therefore provide an equally good reproduction of the data. Using $a=2$ MeV, as in \cite{Saue11}, leads to a change which is too strong when compared to the data. As in \cite{raus12,nett13}, we find that instead using $a=5$ MeV leads to an overall good description of the experimental results. It has to be cautioned, however, that it is not possible to draw a straightforward conclusion due to the additional sensitivity to further widths as discussed above. A modification of all three widths leads to an equal result but yielding a different value for $a$ or even no need for an energy-dependent \a-nucleus potential. Without additional constraints, e.g., (\a,\g) cross sections or further data constraining the neutron- or \g widths, it is impossible to decide which solution is correct.

The above also applies to the recent suggestion of an additional channel contributing to the total cross section but not yet considered in the present Hauser-Feshbach models \cite{Raus13}. Instead of modifying the energy-dependence of the optical potential, leading to a different total cross section, the new idea is that this total cross section is actually described well by the standard potential but it has to be distributed differently between the possible reaction channels. As the current data for $^{166}$Er(\a,n) and $^{165}$Ho(\a,n) do not allow to uniquely determine what modification of the \a-nucleus potential is required, if any, they also do not constrain a possible additional reaction channel.

\section{Summary and conclusion}

The cross sections of the reactions \ho and \er have been measured at energies between $10.5$ MeV and $15$ MeV. The experimental values were determined by the activation technique making use of \g-ray as well as X-ray spectroscopy.

Due to the dominant sensitivity of the reaction cross sections to the \a widths at the upper end of the investigated energy range, it was found that the McFadden and Satchler potential \cite{McFa66} allows the best description of the data at these energies. Deviations between measurement and predictions towards lower energy require modification of the predicted \g-, neutron-, or \a width but an (\a,n) measurement alone does not allow to unambiguously decide which widths have to be changed and in which manner.
Further measurements are required to better constrain all the widths and specifically the \a width at low, astrophysically important energies. Especially, a combination of low-energy ($\alpha$,n) and ($\alpha$,$\gamma$) data on the same target nucleus, supplemented by ($\alpha$,$\gamma$) cross sections below the neutron threshold, would allow to study the energy dependence of the $\alpha$ width towards low energy (see, \eg, \cite{gyur06,kisslett,raus12}). Measurements studying partial cross sections for $\alpha$ emission from the compound nucleus would also probe the energy dependence of the $\alpha$ optical potential (\eg , \cite{raus13,Raus11,raus14a,koeh2001} and references therein) but would require reactions on unstable targets for the cases discussed here.

\begin{acknowledgments}
The authors would like to thank all members of the Nuclear Science Laboratory at University of Notre Dame for their tremendous support. Furthermore, we are grateful to N. Pietralla for providing target materials as well as equipment for spectroscopy. This work is supported by the Deutsche Forschungsgemeinschaft under the contracts SO 907/2-1 and SFB 634 as well as by DAAD (PPP USA, 50141751). Additionally, the project is supported by the HIC for FAIR within the framework of LOEWE launched by the State of Hesse, Germany, by the Swiss NSF, the European Research Council, and the THEXO collaboration within the 7$^{th}$ Framework Programme of the EU.\\
\end{acknowledgments}

\end{document}